\documentclass[manuscript]{aastex62}
\usepackage[encapsulated]{CJK}
\usepackage{subfigure}
\usepackage{mathrsfs}
\usepackage{amsmath}
\usepackage{epstopdf}
\usepackage{longtable}
\usepackage{booktabs}
\usepackage{tablefootnote}
\usepackage{threeparttable}
\usepackage{multirow}
\usepackage{hyperref} 
\def\etal {et al.~}

\newbox\grsign \setbox\grsign=\hbox{$>$} \newdimen\grdimen \grdimen=\ht\grsign
\newbox\laxbox \newbox\gaxbox
\setbox\gaxbox=\hbox{\raise.5ex\hbox{$>$}\llap
	{\lower.5ex\hbox{$\sim$}}}\ht1=\grdimen\dp1=0pt
\setbox\laxbox=\hbox{\raise.5ex\hbox{$<$}\llap
	{\lower.5ex\hbox{$\sim$}}}\ht2=\grdimen\dp2=0pt

\shorttitle{Outflow in Nebula}
\shortauthors{Chen \etal}

\begin{document}
	\begin{CJK*}{UTF8}{gbsn}

		\title{Outflowing Gas in the Planetary Nebula K~3-54}
		
		\correspondingauthor{Xiuhui Chen}
		\email{chenxh@huas.edu.cn}
		
		\author{Xiuhui Chen}
		\affiliation{College of Mathematics and Physics, Hunan University of Arts and Science, Changde 415000, China}
		
		\author{Dejian Liu}
		\affiliation{College of Science, China Three Gorges University, Yichang 443002, China}
		
		\author{Ping Yan}
		\affiliation{Purple Mountain Observatory, Chinese Academy of Sciences, Nanjing 210008, China}
		
		\author{Binggang Ju}
		\affiliation{Purple Mountain Observatory, Chinese Academy of Sciences, Nanjing 210008, China}
		
		\author{Dengrong Lu}
		\affiliation{Purple Mountain Observatory, Chinese Academy of Sciences, Nanjing 210008, China}
		
		\author{Yingjie Li}
		\affiliation{Purple Mountain Observatory, Chinese Academy of Sciences, Nanjing 210008, China}
		
		\begin{abstract}
			As planetary nebulae (PNe) evolve, they develop slow and strong dust-driven stellar winds, making the joint study of dust and gas essential for understanding their nature. As a pilot investigation, we selected PN K~3-54 as our target, the only known PN in the Milky Way to exhibit infrared emission from both graphene (C$_{24}$) and fullerene (C$_{60}$). The gas is traced via molecular line emissions from $^{12}$CO, $^{13}$CO, and C$^{18}$O ($J = 1 \rightarrow 0$), observed using the 13.7 m telescope of the Purple Mountain Observatory. We investigate the dynamics of this PN and identify a bipolar outflow. Preliminary results suggest that the large dynamical timescale of the outflow and the weak shock environment may account for the simultaneous survival of C$_{24}$ and C$_{60}$ within and around PN~K~3-54. 
		\end{abstract}
		

\section{Introduction}

Most low- to intermediate-mass stars (i.e., those with masses of 1--8 $M_{\odot}$) evolve from the main sequence through the asymptotic giant branch (AGB) phase, during which they produce dust-rich stellar winds \citep{Castro-Carrizo+2010, Hofner+2018}. Following the AGB phase, stars transition to the protoplanetary nebula (PPN) stage and eventually evolve into planetary nebula (PN). These stages lead to the formation of a wide variety of nonspherical structures, including elliptical, bipolar, and multipolar geometries \citep[e.g.,][]{Balick+2002, Sahai+2007}. During the transition from the AGB to PPN stage, stars also exhibit high-velocity outflows, with typical expansion velocities exceeding 50--100 km s$^{-1}$ \citep{Bujarraba+2001, Sanchez+2012}.

A PN represents an evolutionary stage in which slow, intense, dust-driven winds (a hallmark of the later stages of AGB evolution) become ionized \citep{Garcia+2012}. Immediately following the AGB phase, dust undergoes numerous physical and chemical processes within PNe, including the formation of fullerene C$_{60}$ \citep{Zhang+2011}; the most widely accepted mechanisms for fullerene formation include the photochemical processing of large polycyclic aromatic hydrocarbons \citep{Berne-Tielens2012} or the photochemical processing or destruction of hydrogenated amorphous carbon grains \citep{Garcia+2010, Gomez-Munoz+2024}. As a relatively stable molecule, fullerene may remain intact (almost) indefinitely in space, potentially playing an important role in circumstellar and/or interstellar chemistry and physics. Indeed, \citet{Garcia+2012} detected infrared (IR) fullerene (C$_{60}$) emission from 16 out of 263 PNe (a detection rate of $\sim$ 6\%), as well as IR graphene (C$_{24}$) emission in three PNe. PN~K~3-54 is the only PN in the Milky Way (MW) to exhibit both C$_{60}$ and C$_{24}$ emissions, a peculiarity that motivated our investigation into the dynamics of this object.

Given that PNe develop slow and strong dust-driven winds, and that dust coexists with gas, we, for the first time, use molecular gas to explore the dynamics of PN~K~3-54. This approach aims to investigate the dynamical environments required for the survival of fullerene (C$_{60}$) and graphene (C$_{24}$) within a PN. CO spectral lines are the most commonly used tracers to reveal outflow activity in star-forming regions \citep[e.g.,][]{Shu+1987, Li+2018}, and have been employed to estimate the mass and momentum of outflowing gas entrained in the stellar winds of protostars \citep[e.g.,][]{Bachiller1996}. CO line emission has also become an important tool for studying dynamics of stellar-mass objects at various evolutionary stages, including the AGB and PN phases \citep[e.g.][]{Bujarraba+2001, Castro-Carrizo+2010, Alonso-Hernandez+2024}. For instance, \citet{Alonso-Hernandez+2024} used CO line emission to calculate the mass loss from a sample of AGB stars; \citet{Bujarraba+2001} determined the mass, momentum, and kinetic energy of the bipolar winds in a sample of PPNe; and \citet{Lorenzo+2021} estimated the mass loss in a sample of PPNe and young PNe.

In this pilot study, we focus on the unique PN~K~3-54, which exhibits IR emission from both C$_{60}$ and C$_{24}$, to investigate its dynamics using CO line emission. The remainder of the paper is organized as follows. Section \ref{sec-data} describes the observations used in this work. Section \ref{sec-res} presents the results, including the identification of a bipolar outflow in PN~K~3-54. Section \ref{sec-dis} discusses the distance and physical properties of PN~K~3-54. Finally, Section \ref{sec-conclusion} gives a summary.

\section{Observations}\label{sec-data}

From 2022 May to December, we observed PN~K~3-54 using the 13.7 m telescope of the Purple Mountain Observatory (PMO 13.7 m telescope, hereafter). Three molecular lines were simultaneously observed: $^{12}$CO ($J = 1 \rightarrow 0$) (115.271 GHz) with a half-power beam width (HPBW) of $\sim$ 50$''$, $^{13}$CO ($J = 1 \rightarrow 0$) (110.201 GHz), and C$^{18}$O ($J = 1 \rightarrow 0$) (109.782 GHz) with an HPBW of $\sim52''$. These molecular lines were observed using the nine-beam Superconducting Spectroscopic Array Receiver system (SSAR) in the sideband separation mode \citep{Shan+2012}, with $^{12}$CO in the upper sideband and $^{13}$CO and C$^{18}$O in the lower sideband. The system temperatures were 250--300 K for $^{12}$CO and 150--200 K for $^{13}$CO and C$^{18}$O. Observations were conducted in the on-the-fly (OTF) mode, and the OTF raw data were gridded into a FITS cube with a pixel size of 30$''$ (each pixel contains a spectrum) using the GILDAS software package \citep{Pety2005}. Each fast Fourier transform spectrometer, with a bandwidth of 1 GHz, provided 16,384 channels, resulting in a spectral resolution of 61 kHz. The corresponding velocity resolution was $\sim$0.16 km s$^{-1}$ for $^{12}$CO and $\sim$0.17 km s$^{-1}$ for $^{13}$CO and C$^{18}$O. All results presented in this work are expressed as brightness temperatures, $T_{\mathrm{b}}$. The total size of the map is $\sim$20$'$ with a total integration time of $\sim$3 hr. 

\section{Results}\label{sec-res}

This section describes the channel map and the identification of a bipolar outflow in PN~K~3-54.

\subsection{Channel Map of PN~K~3-54}\label{sec-channel-map}

\begin{figure*}
	\centering
	\includegraphics[width=0.99\linewidth]{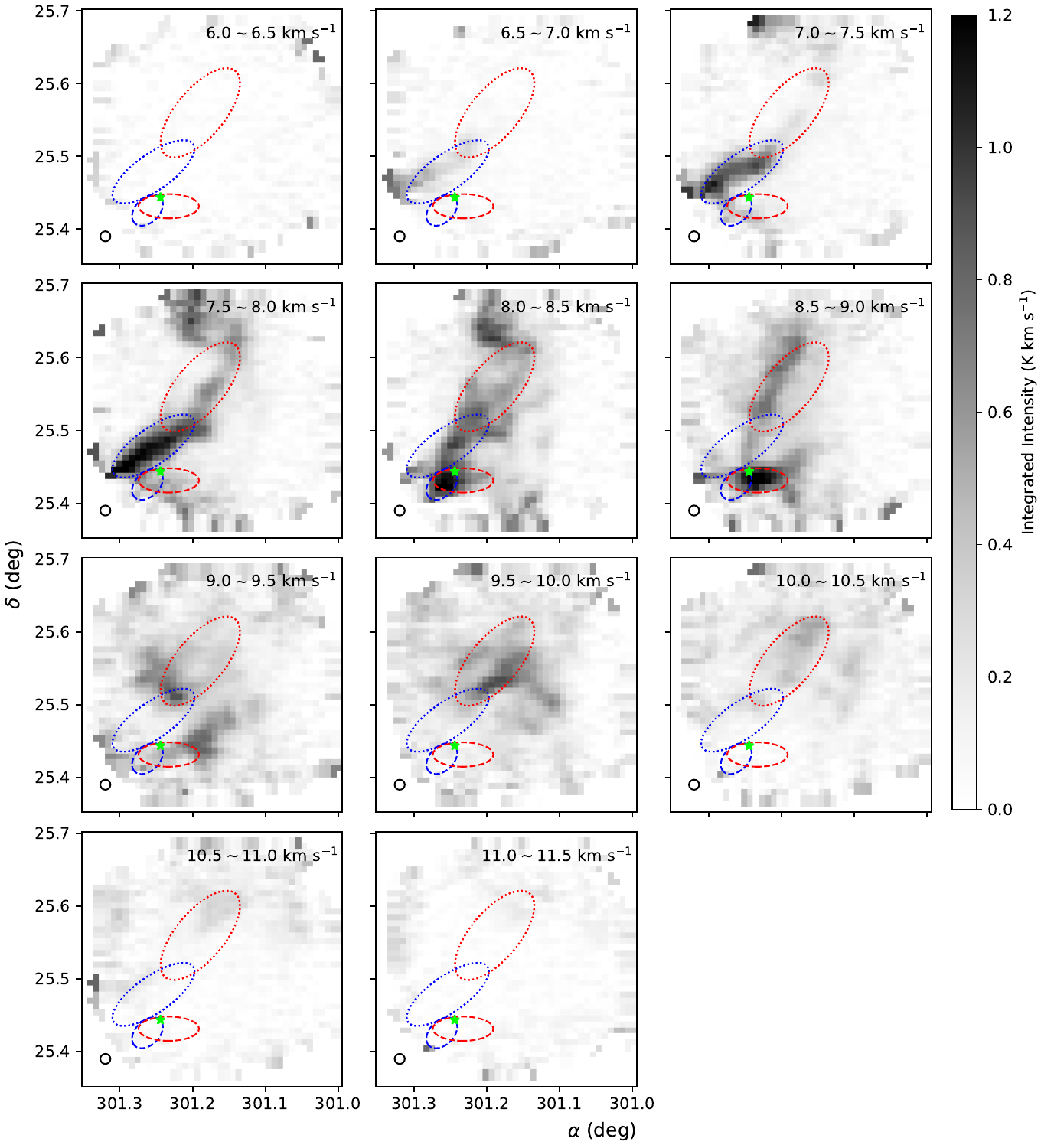}
	\caption{PN~K~3-54 $^{13}$CO channel map, where $\alpha$ and $\delta$ are the R.A. and Decl. (J2000), respectively. The black open circle represents the beam size of $^{13}$CO, and the lime star denotes PN~K~3-54. The dotted ellipses outline a pronounced bipolar structure (see Figure \ref{fig:outflow}), and the dashed ellipses indicate a possible bipolar outflow that requires further confirmation. }
	\label{fig:channel-map}
\end{figure*}

Two components are clearly visible in the channel map shown in Figure \ref{fig:channel-map}. The velocity watershed of these two components is at $\sim$8.5--9.0 km s$^{-1}$. This velocity watershed corresponds to the systemic velocity, which is more precisely determined to be $\sim$8.6 km s$^{-1}$, as shown in Figure \ref{fig:outflow}(c)-(d). The blue (low-velocity) component (see the dotted blue ellipse) begins to appear at a velocity of $\sim$6.5 km s$^{-1}$, reaching its peak emission at $\sim$7.5--8.0 km s$^{-1}$, and fades out at $\sim$8.5 km s$^{-1}$. For the red (high-velocity) component (see the dotted red ellipse), the corresponding velocities are $\sim$9.0, $\sim$9.5--10, and $\sim$11.0 km s$^{-1}$, respectively. The figure exhibits a pronounced bipolar structure with a southeast-northwest orientation, which we identify as a bipolar outflow in PN K 3-54 (see below).

In addition, another bipolar elongated structure with a very small velocity span is visible in Figure \ref{fig:channel-map}. This structure consists of a low-velocity component at $\sim$8.0--8.5 km s$^{-1}$ (see the dashed blue ellipse in Figure \ref{fig:channel-map}) and a high-velocity component at $\sim$9.0--9.5 km s$^{-1}$ (see the dashed red ellipse). The blue elongated structure is located to the southeast of PN~K~3-54 (denoted by a lime star) and is aligned along a southeast--northwest direction, while the red elongated structure, located to the west of the blue structure, exhibits an east--west orientation.
This possible bipolar structure may trace a slow outflow roughly along the east-west direction. If confirmed by three-dimensional maps with high spatial and velocity resolution, the presence of two bipolar outflows with different orientations may suggest that the orientation of the stellar winds driven by the central star varies over time.

\subsection{Outflow Identification}\label{sec-outflow}

An outflow can be inferred and characterized based on three aspects: the spectral line profiles, the spatial distribution of the blue and red lobes (traced by the integrated intensity map of each lobe), and the position--velocity (P-V) diagram. PNe exhibit a bimodal line profile in CO emission, as shown by \citet{Castro-Carrizo+2010} and \citet{Lorenzo+2021}. This bimodal line profile suggests that a large number of ejections have accumulated around PNe, with the broad line wings serving as tracers of these ejections. Therefore, throughout this work, regions corresponding to the bimodal line profile are treated as outflowing gas components.

\begin{figure*}
	\centering
	\subfigure[Spatial distribution of outflow]{\includegraphics[height=0.26\textheight]{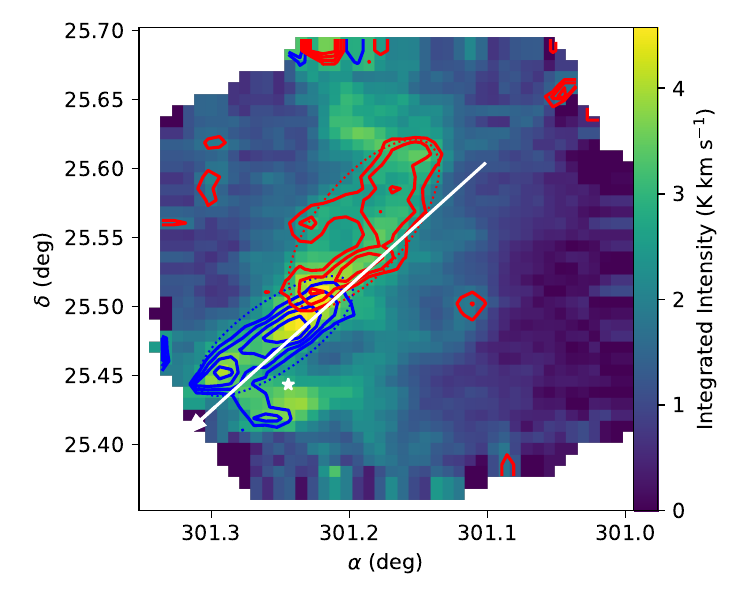}}
	\subfigure[P-V diagram]
	{\includegraphics[width=0.3\textheight]{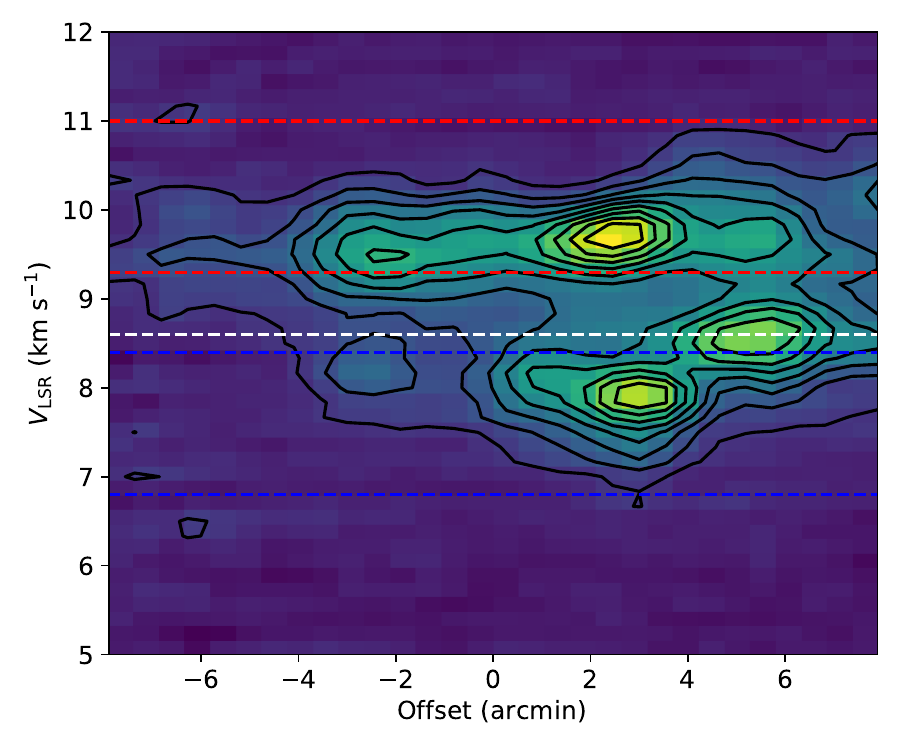}}
	\subfigure[Spectrum of the blue lobe]
	{\includegraphics[width=0.35\textheight]{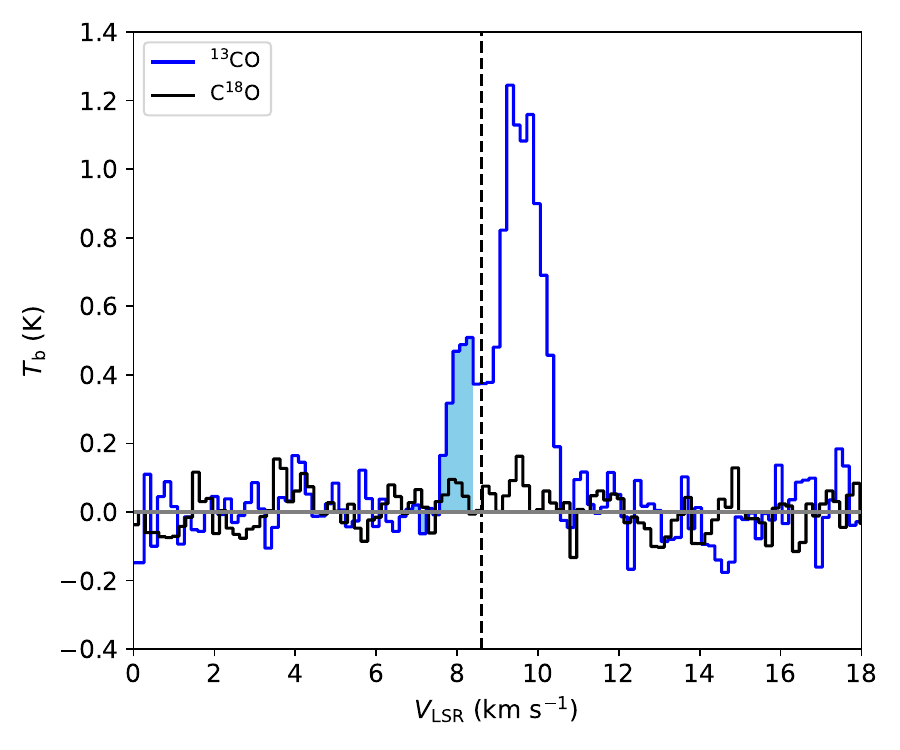}}
	\subfigure[Spectrum of the red lobe]
	{\includegraphics[width=0.35\textheight]{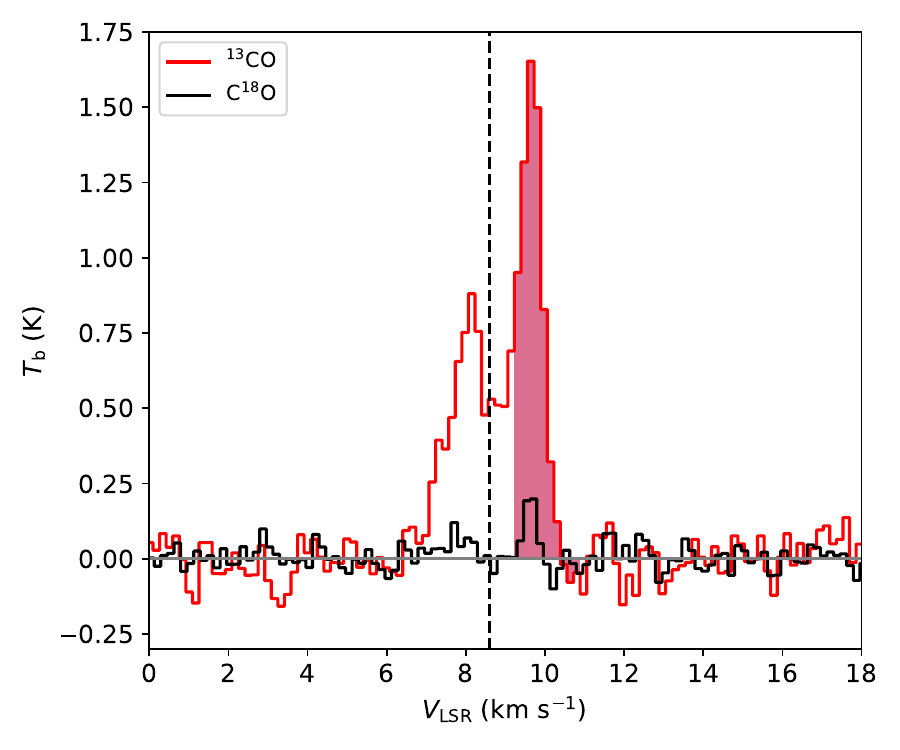}}
	\caption{Outflowing gas in PN~K~3-54 identified by $^{13}$CO emission. (a) Integrated C$^{18}$O map with $^{13}$CO blue and red lobe contours overplotted. The contour levels range from 60\% to 90\% with steps of 10\% of the peak intensity of each lobe. The lengths of the major axes of the ellipses, representing the red and blue blue lobes, are used to estimate the length of outflow lobes. The white star represents PN~K~3-54, and $\alpha$ and $\delta$ refer to R.A. and Decl. (J2000), respectively. (b) P-V diagram along the white arrow in panel (a). The black contour levels range from 10\% to 90\% with steps of 10\% of the peak value. The blue and red dashed lines show the velocity intervals of the blue and red lobes, respectively. The white dashed line indicates the systemic velocity of PN K 3-54. (c) Spectra of $^{13}$CO and C$^{18}$O, shown in blue and black, respectively, at the pixel corresponding to the peak position of the blue emission of the $^{13}$CO outflow. The blue shading in the spectrum indicates the blue line wing velocity of $^{13}$CO, and the black dash line shows the systemic velocity. (d) The spectra of $^{13}$CO and C$^{18}$O, shown in red and black, respectively, at the pixel corresponding to the peak position of the red emission of the $^{13}$CO outflow. The red shading in the spectrum indicates the red line wing velocity of $^{13}$CO, and the black dash line shows the systemic velocity.}
	\label{fig:outflow}
\end{figure*}

Figure \ref{fig:outflow} shows a clear bipolar outflow traced by $^{13}$CO emission in PN~K~3-54. Panel (a) presents an integrated intensity map of C$^{18}$O with the $^{13}$CO blue and red lobe contours overplotted. The integrated intensity map of C$^{18}$O shows the overall distribution of gas across the region. The red and blue lobes of $^{13}$CO (see the red and blue dotted ellipses, respectively) demonstrate relatively symmetric, clearly elongated structures with high collimation (i.e., the major/minor axis radii of the red and blue ellipses are $\sim$2.3$'$ and 2.7$'$, respectively). Panel (b) shows a P-V diagram along the direction connecting the peak emission from the red and blue lobes (see the white arrows in panel (a)). Panels (c) and (d) show velocity profiles and selected velocity intervals of the line wings (see the shaded regions), where spectra of the blue and red lobes are extracted from pixels corresponding to the peak integrated intensity of the respective lobes. Given that the line wings are relatively faint, we selected the brightest pixels to more effectively capture the spectral line profiles of the line wings. The inspection of the entire figure reveals a clear bipolar outflow structure in PN~K~3-54, evident in the spectral line profiles, spatial distributions of the blue and red lobes, and the P-V diagram. The bimodal line profiles shown in panels (c) and (d) are characteristic of PNe \citep[e.g.,][]{Castro-Carrizo+2010, Lorenzo+2021}, suggesting that the excitation source for this bipolar outflow is probably the central source of PN~K~3-54.

\section{Discussion}\label{sec-dis}

\subsection{Estimation of Distance}\label{subsec-est}

The distance to PN~K~3-54 was assumed to be 23.4 kpc by \citet{Stanghellini-Haywood2010} and \citet{Garcia+2012}. The parallax from Gaia Data Release 3 (DR3) of the corresponding star is -0.04 $\pm$ 0.13 mas \citep[the Gaia DR3 ID: 1834720521736114816;][]{Gaia-Collaboration+2021}, indicating that no effective distance is provided. For stars with a poorly determined or unknown Gaia parallax, \citet{Bailer-Jones+2021} attempted to determine stellar distances by combining the colors and apparent magnitudes of stars with known parallaxes. This approach assumes that a star with a given color has a limited number of possible absolute magnitudes (plus extinction). For the star of interest in this work, \citet{Bailer-Jones+2021} derived a median distance of 4.087$^{+0.906}_{-1.718}$ kpc. However, the actual degree of extinction toward PN~K~3-54 affects the reliability of the distance inferred by this method. \citet{Nataf+2016} argued that the extinction law varies across the inner Galaxy, and the Galactic coordinates of PN~K~3-54 are $l \sim$63.8$^{\circ}$ and $b \sim$$-$3.3$^{\circ}$, respectively.

\begin{figure*}[!htb]
	\centering
        \includegraphics[width=0.7\textwidth]{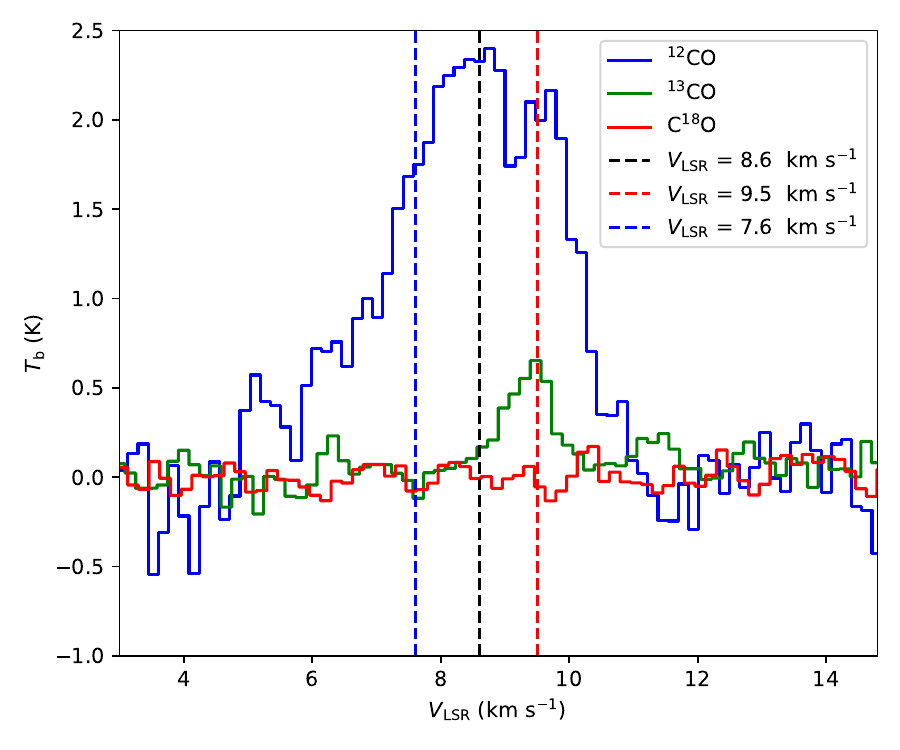}
	\caption{Spectra centered on PN~K~3-54, where the blue, green and red lines represent the spectra of $^{12}$CO, $^{13}$CO, and C$^{18}$O, respectively. The rms values for these three spectra are 0.21, 0.09, and 0.08 K, respectively. The black, red, and blue vertical lines indicate the systemic velocity and the velocities at the peak emission of the red and blue lobes, respectively. The threshold for a confirmed detection is that the values of $T_{\mathrm{b}}$ in three consecutive channels (corresponding to $\sim$0.5 km s$^{-1}$) exceed 3$\times$rms. Although no C$^{18}$O emission was detected, the C$^{18}$O spectrum is included here for reference, providing further context for this region.}
	\label{fig:Kinematic distance}
\end{figure*}

The kinematic distance to PN~K~3-54 was determined using the model developed by \citet{Reid+2016, Reid+2019}, which applies a Bayesian approach to distance estimation. This method integrates the MW's rotation curve, Galactic coordinates, radial velocity, and other available data, such as proper motion. \footnote{This approach constructs a probability density function (PDF) for each type of available distance information, and these PDFs are subsequently multiplied to obtain a combined distance PDF. The final distance is determined from the combined distance PDF.}The key parameters for estimating the kinematic distance are the Galactic coordinates, and the radial velocity, $V_{\mathrm{LSR}}$, of the target source. Figure \ref{fig:Kinematic distance} presents the spectra of PN~K~3-54. The red line wing (corresponding to the red lobe of the bipolar outflow in Figure \ref{fig:outflow}) is centered at $\sim$9.5 km s$^{-1}$, as indicated in both the spectrum from $^{12}$CO (Figure \ref{fig:Kinematic distance}) and the $^{13}$CO spectra at the peak emission of the blue and red lobes (see panels (c) and (d) of Figure \ref{fig:outflow}) of PN~K~3-54. The blue line wing (corresponding to the blue lobe of the bipolar outflow in Figure \ref{fig:outflow}) is blended with a core component in the $^{12}$CO spectrum and obscured by noise in the $^{13}$CO spectrum (Figure \ref{fig:Kinematic distance}). Fortunately, the blue line wing is visible at other positions (e.g., panels (c) and (d) in Figure \ref{fig:outflow}) with a central velocity of $\sim$7.6 km s$^{-1}$. The central velocity of the core component (i.e, the systemic velocity) is pronounced at $\sim$8.6 km s$^{-1}$ when combining the $^{13}$CO spectra in panels (c) and (d) of Figure \ref{fig:outflow} with the $^{12}$CO spectrum in Figure \ref{fig:Kinematic distance}. Adopting the values of $l$, $b$ and $V_{\mathrm{LSR}}$ as $63.80^{\circ}$, $-3.30^{\circ}$, and 8.6 km s$^{-1}$, respectively, the corresponding kinematic distance estimate,\footnote{See \url{http://bessel.vlbi-astrometry.org/node/378}} following \citet{Reid+2019}, is roughly $\sim$0.4 $\pm$ 0.8 kpc. This distance of 0.4 kpc is adopted as the final distance to PN~K~3-54 (see below).

\subsection{Physical Properties}\label{subsec-cal}

The physical properties of the bipolar outflow in PN~K~3-54, traced by $^{13}$CO line emission, were calculated assuming a distance of 0.4 kpc. Under the assumption of local thermodynamic equilibrium (LTE), the H$_2$ column density, $N_{\mathrm{lobe}}$, of the outflow lobe is derived from \citep{Wilson+2013}
\begin{equation}\label{equ:N}
	N_{\mathrm{lobe}} = \frac{4.6 \times 10^{13}}{X(^{13}\mathrm{CO})} \frac{T_{\mathrm{ex}}}{e^{-5.3/T_{\mathrm{ex}}}} \int T_{\mathrm{b}} dv,
\end{equation}
where the velocity range corresponds to the line wing interval (see Figure \ref{fig:outflow}). The abundance of $^{13}$CO is assumed to be $X(^{13}\mathrm{CO}) = [^{13}\mathrm{CO}]/[\mathrm{H_2}] = 2 \times 10^{-5}$ \citep{Bujarraba+2001}. The excitation temperature, $T_{\mathrm{ex}}$, is taken as the dust temperature determined by Planck \citep{Planck-Collaboration+2011, Planck-Collaboration+2016} in this region,  which is 18 K, similar to the value used by \citet{Bujarraba+2001}. The relative uncertainty in $N_{\mathrm{lobe}}$ due to $T_{\mathrm{ex}}$ is less than 50\%, assuming $T_{\mathrm{ex}}$ ranges from 10 to 30 K. Because there is only one component of $^{12}$CO emission (with a systemic velocity of $\sim$8.6 km s$^{-1}$) within the velocity range of [-100, 100] km s$^{-1}$, dust is probably associated with this component. 
The mass of the outflow lobe is
\begin{equation}\label{equ:M}
	M_{\mathrm{lobe}} = N_{\mathrm{lobe}} A_{\mathrm{lobe}} \mu m_{\mathrm{H}},
\end{equation}
where $A_{\mathrm{lobe}}$ is the area enclosing 60\% of the peak of outflow lobe, $\mu = 2.72$ is the average molecular mass, and $m_{\mathrm{H}}$ is the mass of the hydrogen atom \citep{Garden+1991}.

We also calculated the momentum, $P_{\mathrm{lobe}}$, and kinetic energy, $E_{\mathrm{lobe}}$, of the outflow lobe, as well as the average velocity (i.e., the relative velocity with respect to the systemic system, weighted by $T_{\mathrm{b}}$), $\langle \Delta v_{\mathrm{lobe}}\rangle$,
\begin{equation}\label{equ:P}
	P_{\mathrm{lobe}} = \sum_{A_{\mathrm{lobe}}}M_{\mathrm{lobe}} \langle \Delta v_{\mathrm{lobe}}\rangle,
\end{equation}
\begin{equation}\label{equ:E}
	E_{\mathrm{lobe}} = \sum_{A_{\mathrm{lobe}}}M_{\mathrm{lobe}} \langle \Delta v^2_{\mathrm{lobe}}\rangle,
\end{equation}
where $\langle \Delta v^2_{\mathrm{lobe}}\rangle$ represents the square of the average velocity \citep[for further details, see][]{Li+2018, Liu+2021}. Additionally, we calculated the dynamical timescale, $t_{\mathrm{lobe}}$, the mechanical luminosity, $L_{\mathrm{lobe}}$, and the mass-loss rate, $\dot{M}_{\mathrm{lobe}}$, of the outflow lobe:
\begin{equation}\label{equ:t}
	t_{\mathrm{lobe}} = l_{\mathrm{lobe}}/\Delta v_{\mathrm{max}},
\end{equation}
\begin{equation}\label{equ:L}
	L_{\mathrm{lobe}} = E_{\mathrm{lobe}}/t_{\mathrm{lobe}},
\end{equation}
\begin{equation}\label{equ:dot{M}}
	\dot{M}_{\mathrm{lobe}} = M_{\mathrm{lobe}}/t_{\mathrm{lobe}},
\end{equation}
where $l_{\mathrm{lobe}}$ is the length of the outflow lobe, which was estimated from the ellipses shown in Figure \ref{fig:outflow}(a), and $\Delta v_{\mathrm{max}}$ is the maximum velocity of the outflow lobe. The values of $\Delta v_{\mathrm{max}}$ for the red and blue lobes are equal to the absolute value of the maximum velocity difference between the velocity indicated by the white line and those of the red and blue lines, respectively, as shown in Figure \ref{fig:outflow}(b).

Assuming a random distribution of inclination angles, the mean value of inclination angle is given by $\langle \theta \rangle = \int_{0}^{\pi/2} \theta \cos \theta d\theta \sim 33^{\circ}$ \citep[i.e., the same as that adopted in][where the inclination angle is not known]{Bujarraba+2001}. The corrections for an inclination angle of $33^{\circ}$ to the values of $\langle \Delta v_{\mathrm{lobe}}\rangle$, $l_{\mathrm{lobe}}$, $P_{\mathrm{lobe}}$, $E_{\mathrm{lobe}}$, $t_{\mathrm{lobe}}$, $L_{\mathrm{lobe}}$, and $\dot{M}_{\mathrm{lobe}}$ are $1/\sin \theta = 1.8$, $1/\cos \theta = 1.2$, $1/\sin \theta = 1.8$, $1/\sin^2 \theta = 3.4$, $\sin \theta/\cos \theta = 0.6$, $\cos \theta/\sin^3 \theta = 5.2$, and $\cos \theta/\sin \theta = 1.5$, respectively. On the other hand, if the true inclination of the outflow with respect to the plane of the sky is close to 0$^{\circ}$, we may significantly underestimate $\langle \Delta v_{\mathrm{lobe}}\rangle$, $P_{\mathrm{lobe}}$, $E_{\mathrm{lobe}}$, $L_{\mathrm{lobe}}$, and $\dot{M}_{\mathrm{lobe}}$. The probability that $\theta$ is smaller than 20$^{\circ}$, assuming a random distribution of the outflow axes, is $\sim$0.40. Therefore, with a probability of $\sim$60\%,  the underestimates for the values of $\langle \Delta v_{\mathrm{lobe}}\rangle$, $P_{\mathrm{lobe}}$, $E_{\mathrm{lobe}}$, $L_{\mathrm{lobe}}$, and $\dot{M}_{\mathrm{lobe}}$ are smaller by factors of 1.6, 1.6, 2.5, 4.5, and 1.8, respectively. However, for the values of $l_{\mathrm{lobe}}$ and $t_{\mathrm{lobe}}$, the corrections for inclination angles of $20^{\circ}$ (which are 1.1 and 0.4, respectively) are smaller than those for inclination angles of $33^{\circ}$.

Table \ref{tab:outflow-parameter} lists the physical properties of the outflow lobe, corrected for the inclination angle (assuming $\theta = 33^{\circ}$), with $\langle \Delta v_{\mathrm{lobe}}\rangle$ as the only physical parameter independent of distance. Note that the distance is embedded in the units of the results presented in the table due to the large uncertainty in the distance. If the distance to PN~K~3-54 is 23.4 kpc, the value of $l_{\mathrm{lobe}}$ would be $\sim$7 pc, which exceeds the maximum size a PN could eventually reach \citep[see][]{Dopita+1991, Pierce+2004, Parker2022}, suggesting that this distance is likely an overestimate. With a distance of $\sim$4 kpc, the sum of mass of the blue and red lobes would be $\sim$218 $M_{\odot}$, which far exceeds the mass of the outflows observed toward PPNe as reported by \citet{Bujarraba+2001}. Similarly, the sum of the mass-loss rates of the blue and red lobes would reach 1.1 $\times$ 10$^{-3}$ $M_{\odot}$ yr$^{-1}$, a value that is also much higher than those measured for AGB and early post-AGB circumstellar envelopes \citep{Castro-Carrizo+2010}, and for PPNe and young PNe \citep{Lorenzo+2021}. These values suggest that the distance of $\sim$4 kpc is also likely an overestimate. Therefore, the adopted distance to PN~K~3-54 of $\sim$0.4 kpc seems more reasonable. Note, however, the actual distance requires further confirmation.

\begin{table*}[htbp]
	\scriptsize
	\setlength\tabcolsep{1.5pt}
	\caption{Physical Properties of the Outflow PN~K~3-54 with Inclination Correction \label{tab:outflow-parameter}}
	\begin{tabular}{lccccccccc}
		\hline \hline			
		Lobe	&	$N_{\mathrm{lobe}}$	&	$M_{\mathrm{lobe}}$	&	$\langle \Delta v_{\mathrm{lobe}}\rangle$	&	$l_{\mathrm{lobe}}$	&	$P_{\mathrm{lobe}}$	&	$E_{\mathrm{lobe}}$	&	$t_{\mathrm{lobe}}$	&	$L_{\mathrm{lobe}}$ & $\dot{M}_{\mathrm{lobe}}$	\\
		&	(10$^{20}$ cm$^{-2}$)	&	$\left(\displaystyle \frac{d^2}{0.16} M_{\odot}\right)$	&	(km s$^{-1}$)	&	$\left(\displaystyle\frac{d}{0.4} \mathrm{pc} \right)$	&	$\displaystyle\left(\frac{d^2}{0.16} M_{\odot}\; \mathrm{km\;s^{-1}}\right)$	&	$\left(\displaystyle \frac{d^2}{0.16} 10^{44} \; \mathrm{erg}\right)$	&	$\left(\displaystyle \frac{d}{0.4} 10^4 \; \mathrm{yr}\right)$	&	$\left(\displaystyle \frac{d}{0.4} L_{\odot}\right)$ & $\left(\displaystyle \frac{d}{0.4} M_{\odot} 10^{-5} \; \mathrm{yr^{-1}}\right)$	\\
		\hline
		blue	&	1.3	&	1.3	&	3.1	&	0.1	&	7.4	&	4.1	&	2.1	&	1.0	& 6.0 \\
		red	&	0.7	&	0.9	&	4.5	&	0.1	&	7.9	&	6.8	&	1.7	&	1.6	& 5.7 \\
		Mean & 2.0 & 1.1 & 3.8 & 0.1 & 7.7 & 5.5 & 1.9 & 1.3 & 3.8 \\
		Sum  & ... & 2.2 & ... & ... & 15.3 & 10.9 & ... & 2.6 & 11.3 \\
		\hline		
	\end{tabular}
	\begin{tablenotes}
		\item{Note. $d$ denotes the actual distance to PN~K~3-54.}
	\end{tablenotes}
\end{table*}

The average dynamical timescale, $t_{\mathrm{lobe}}$, for the two lobes is $\sim$$1.9 \times 10^4$ yr. Even assuming $\theta = 20^{\circ}$, the average dynamical timescale decreases to $\sim$$1.1 \times 10^4$ yr, which is
larger than the dynamical ages of PNe reported in previous studies \citep[i.e., $\sim$400--10000 yr, ][]{Guillen+2013, Danehkar2022, Gomez-Munoz+2023, Derlopa+2024}. The average value of the average velocity, $\langle \Delta v_{\mathrm{lobe}}\rangle$ (independent of distance), of the two lobes is only $\sim$3.8 km s$^{-1}$. Even assuming $\theta = 20^{\circ}$, $\langle \Delta v_{\mathrm{lobe}}\rangle$ increases to only $\sim$6 km $^{-1}$, which is still smaller than the values traced by CO line emission for PPNe, compact PNe, and extended PNe, as reported by \citet{Lorenzo+2021}. This suggests that PN~K~3-54 may be in a relatively weak shock environment. The mass-loss rate, $\dot{M}_{\mathrm{lobe}}$, of PN~K~3-54 is larger than those of AGB and early post-AGB circumstellar envelopes as determined by \citet{Castro-Carrizo+2010} and compact PNe as measured by \citet{Lorenzo+2021}, but it is comparable to the results found for PPNe and extended PNe by \citet{Lorenzo+2021}.

Other physical parameters, such as $N_{\mathrm{lobe}}$, $M_{\mathrm{lobe}}$, $P_{\mathrm{lobe}}$, $E_{\mathrm{lobe}}$, and $L_{\mathrm{lobe}}$, are also listed in Table \ref{tab:outflow-parameter}. The maximum column density, $N_{\mathrm{lobe}}$, for the two lobes is lower than the values reported by \citet{Alonso-Hernandez+2024}, who focused on the envelopes of AGB star (note that they listed column densities of CO). The total value of $M_{\mathrm{lobe}}$ is greater than the majority of the values reported by \citet{Bujarraba+2001}, who focused on PPNe. Meanwhile, total values of $P_{\mathrm{lobe}}$ and $E_{\mathrm{lobe}}$ fall within the ranges of corresponding values found for fast outflows in the study by \citet{Bujarraba+2001}.

In summary, from the perspective of gas dynamics, the outflow velocity $\langle \Delta v_{\mathrm{lobe}}\rangle$ in PN~K~3-54 is relatively smaller compared to that of outflows from AGB stars, PPNe, and PNe, indicating that PN~K~3-54 may be in a weak shock environment. The dynamical timescale may exceed the kinematical ages of PNe. The survival of C$_{24}$ and C$_{60}$ in the environment within and surrounding PN~K~3-54 may be attributed to the large dynamical timescale and the weak shocks present in the environment. However, the conditions under which C$_{24}$ and/or C$_{60}$ formed cannot be determined, as the CO gas studied here is likely not close enough to the central star due to the moderate spatial resolution. In future studies, we will explore the potential correlation between the gas dynamics of PNe and their carbonaceous dust IR emission by conducting a statistical study of CO emission from a large sample of PNe. Such an approach could provide insight into the dynamic environments of carbonaceous dust and assess whether such dust can survive within these conditions. The formation of carbonaceous dust will also be investigated using high-spatial-resolution facilities, such as the Atacama Large Millimeter/submillimeter Array (ALMA).

\section{Summary and Conclusion}\label{sec-conclusion}

Using $^{12}$CO, $^{13}$CO, and C$^{18}$O (J = 1-0) line emission observed with the PMO 13.7 m telescope, we investigated the dynamics of PN~K~3-54, the only PN in the MW where IR emission from both fullerene C$_{60}$ and graphene C$_{24}$ has been detected. A bipolar outflow was identified in this region, and the kinematic distance to PN~K~3-54 was estimated to be roughly $\sim$0.4 kpc. The physical properties of this outflow were calculated, suggesting that the large dynamical timescale of outflow in PN~K~3-54 and the small outflow velocity (indicative of a weak shock environment) may be the reasons why C$_{24}$ and C$_{60}$ can survive simultaneously within and around PN~K~3-54. Further observational studies of molecular gas toward a large number of PNe are required to further confirm our conclusions.

\begin{acknowledgements}
We would like to thank the anonymous referee for the helpful comments and suggestions that helped to improve the paper. We would like to thank all the staff members of the PMO 13.7 m telescopes for their support in making the observations presented in this work.  This work was funded by the Doctoral Research Initiation Fund project of the Hunan University of Arts and Science under grant numbers 19BSQD38.
\end{acknowledgements}

\bibliographystyle{plainnat}
\bibliography{chenxh}
\end{CJK*}
\end{document}